\documentclass[12pt]{amsart}

\usepackage[round]{natbib}
\usepackage{graphicx}
\usepackage{url}
\usepackage{xr}
\usepackage{palatino}
\newcommand{\EE}{\mathbb{E}}

\newcommand{\PD}{\ensuremath{\operatorname{PD}_{\operatorname{u}}}}
\newcommand{\BWPDa}[1]{\ensuremath{\operatorname{BWPD}_{#1}}} % a is for argument.
\newcommand{\qDT}[1]{\ensuremath{^{#1}\!\operatorname{D(T)}}}
\newcommand{\BWPD}{\BWPDa{1}}
\newcommand{\BWPDt}{\BWPDa{\theta}} % t is for theta.
\newcommand{\RBWPDa}[1]{\ensuremath{\operatorname{RBWPD}_{#1}}}
\newcommand{\RBWPDt}{\RBWPDa{\theta}}
\newcommand{\dnPD}{\ensuremath{\Delta\operatorname{nPD}}}
\newcommand{\PDaw}{\ensuremath{\operatorname{PD}_{\operatorname{aw}}}}
\newcommand{\guppy}{\textsf{guppy}}
\newcommand{\pplacer}{\textsf{pplacer}}
\newcommand{\R}{\textsf{R}}
\newcommand{\USEARCH}{\textsf{USEARCH}}
\newcommand{\vegan}{\textsf{vegan}}
\newcommand{\ggplot}{\textsf{ggplot2}}
\newcommand{\boot}{\textsf{boot}}
\newcommand{\Hmisc}{\textsf{Hmisc}}
\newcommand{\picante}{\textsf{picante}}
\newcommand{\PQE}{\operatorname{PQE}}

\newcommand{\arxiv}[1]{#1}
\newcommand{\notarxiv}[1]{}
\newcommand{\eat}[1]{}

\newcommand{\FIGshape}{\
\begin{figure}
\begin{center}
  \includegraphics[height=8cm]{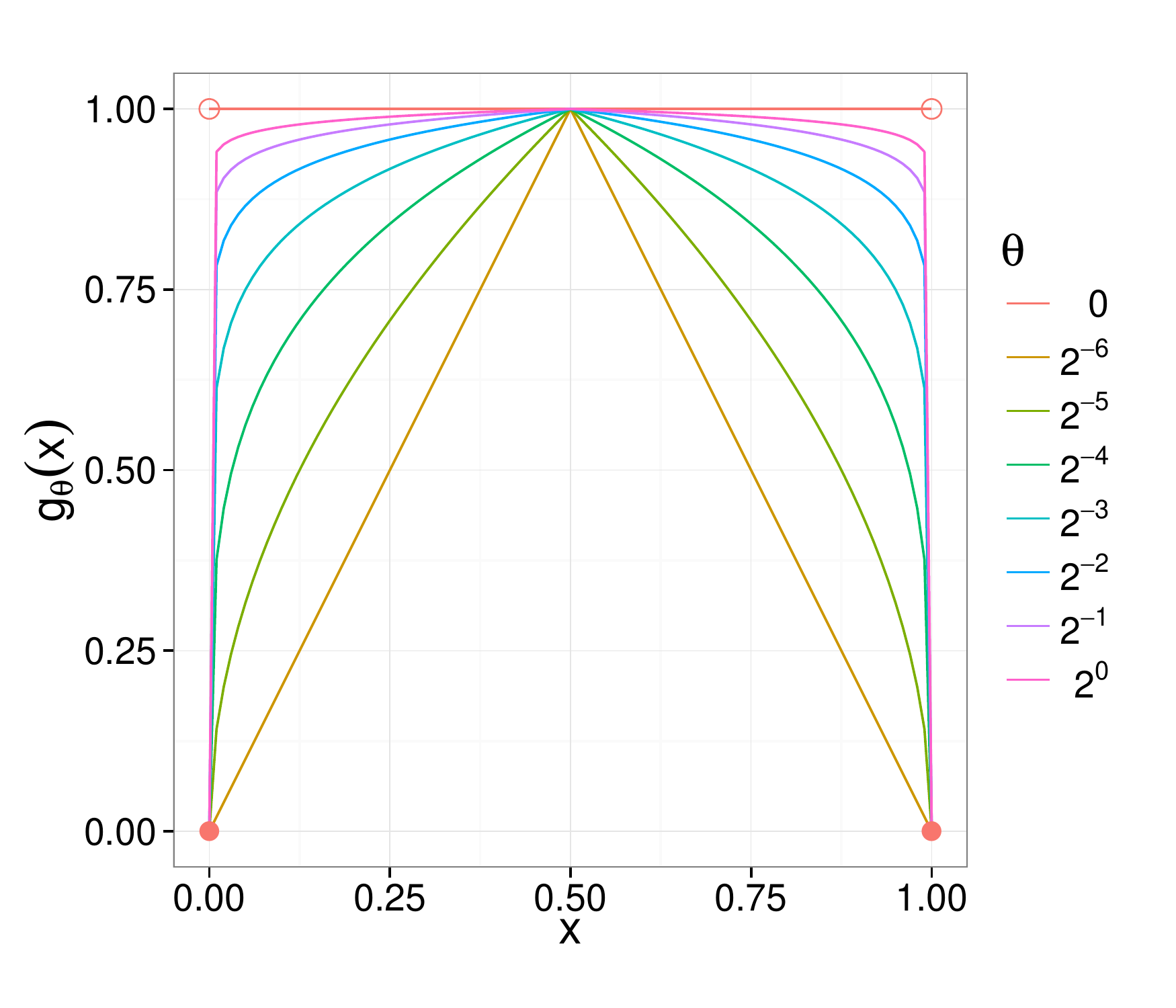}
\end{center}

\caption{\
  $g_\theta$ curves for various $\theta$ parameters.
  As $\theta$ goes to zero, the $g_\theta$ converge pointwise to $g$, which is 1 on the interior of the unit interval and 0 on the boundaries.
}
\label{FIGshape}
\end{figure}
}

\newcommand{\FIGvaginaldendro}{\
  \begin{figure}
    \begin{center}
      \includegraphics[height=7cm]{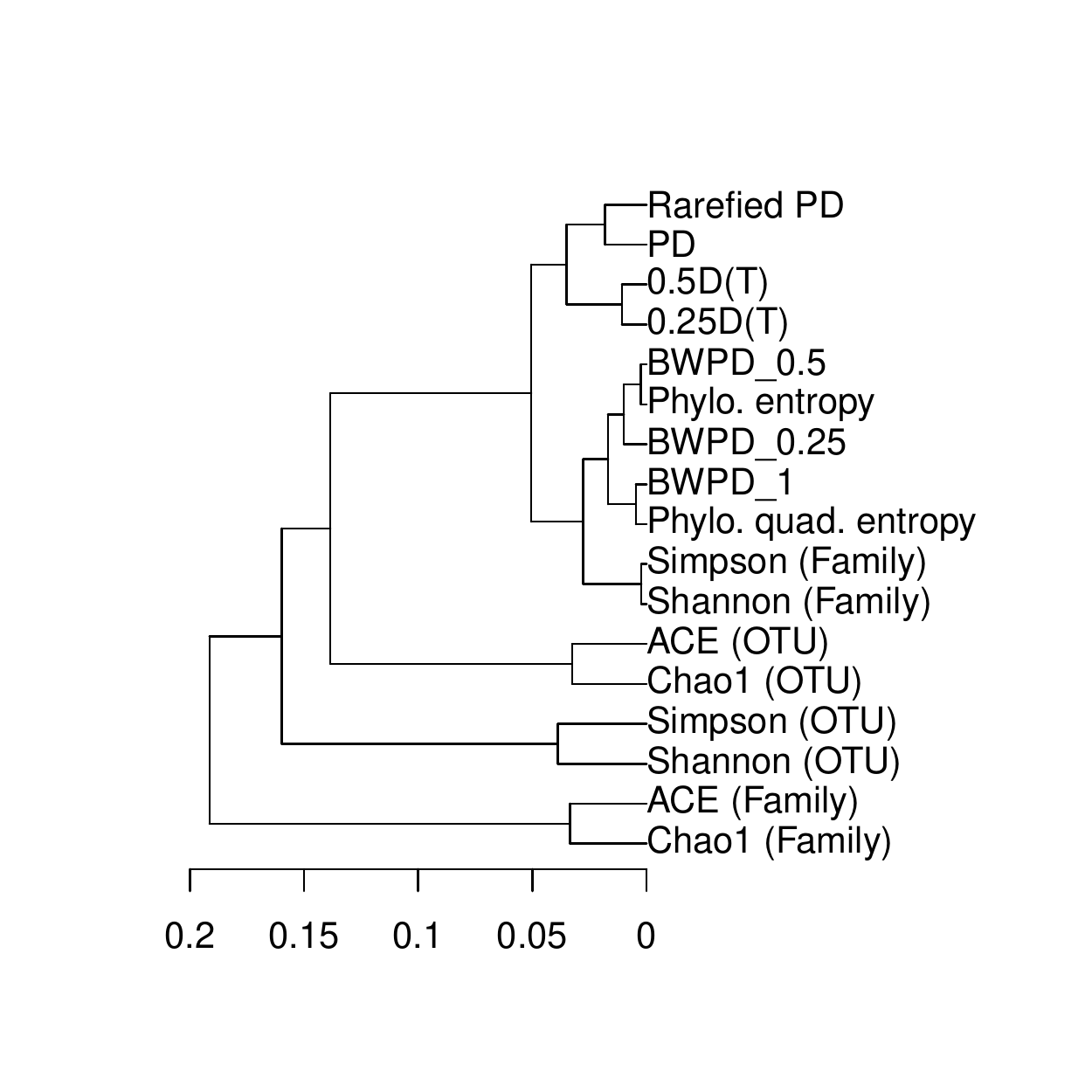}
    \end{center}
    \caption{Dendrogram relating alpha diversity measures applied to the vaginal dataset.}
    \label{FIGvaginaldendro}
  \end{figure}
}

\newcommand{\FIGvaginalrarefy}{\
  \begin{figure}
    \begin{center}
      \centerline{\includegraphics[height=9cm]{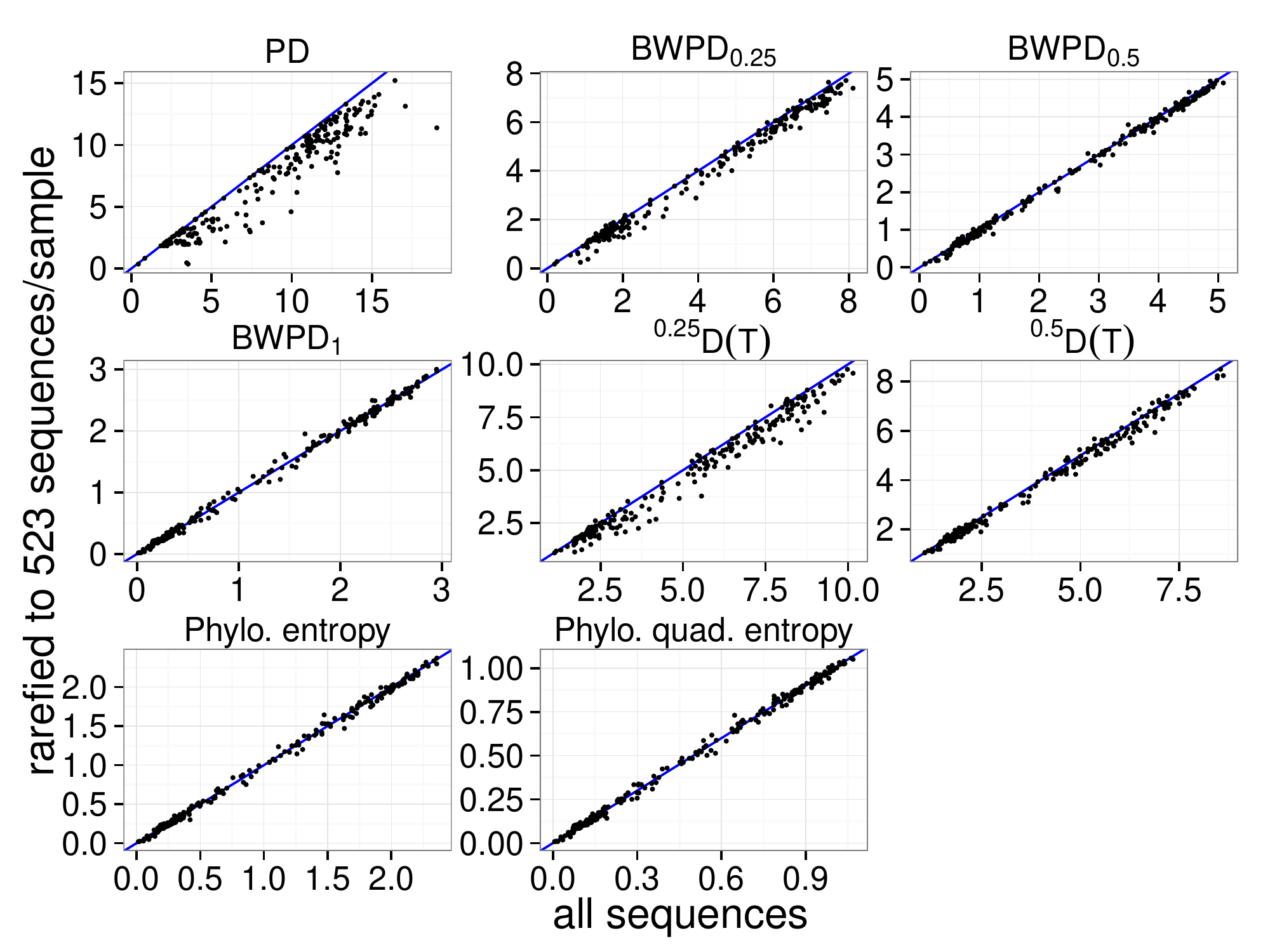}}
    \end{center}
    \caption{Comparison of rarefied and unrarefied values of various phylogenetic alpha diversity measures as applied to the vaginal dataset.
      The value of six alpha measures for each specimen using all available sequences is plotted on the $x$-axis.
      The value of the alpha measures for each specimen after a single rarefaction to 523 sequences (the smallest sequence count across specimens) is plotted on the $y$-axis.
      The $y=x$ line is shown in blue.
    }
    \label{FIGvaginalrarefy}
  \end{figure}
}

\newcommand{\TABLEvaginalcorrelation}{\
\begin{table}[ht]
\centering
\begin{tabular}{lrrr}
  \hline
  Measure & Nugent $R^2$ & Amsel Accuracy & Amsel p-value \\
  \hline
  \BWPDa{0.25}                    & 0.738 &  0.828 &  1.49E-35 \\
  Simpson (Family)                & 0.731 &  0.822 &  2.07E-33 \\
  Rarefied $\PD$                  & 0.731 &  0.828 &  6.81E-35 \\
  Shannon (Family)                & 0.721 &  0.821 &  8.85E-33 \\
  \BWPDa{0.5}                     & 0.703 &  0.823 &  2.16E-33 \\
  $\PD$                           & 0.696 &  0.832 &  1.26E-32 \\
  Phylo. entropy                  & 0.679 &  0.832 &  1.56E-31 \\
  \qDT{0.25}                      & 0.677 &  0.818 &  8.74E-30 \\
  \qDT{0.5}                       & 0.662 &  0.814 &  5.35E-29 \\
  Phylo. quad. entropy            & 0.647 &  0.811 &  7.70E-30 \\
  \BWPDa{1}                       & 0.610 &  0.796 &  5.66E-28 \\
  Chao1 (Family)                  & 0.610 &  0.823 &  9.79E-24 \\
  Chao1 (OTU)                     & 0.450 &  0.758 &  1.61E-19 \\
  ACE (OTU)                       & 0.422 &  0.763 &  6.86E-20 \\
  Shannon (OTU)                   & 0.380 &  0.754 &  6.73E-16 \\
  Simpson (OTU)                   & 0.192 &  0.700 &  1.36E-07 \\
  ACE (Family)                    & 0.088 &  0.666 &  1.51E-01 \\
   \hline
\end{tabular}
\caption{\
  Correlation and predictive performance of the various alpha diversity measures, ordered by decreasing $R^2$ value.
  Nugent $R^2$: $R^2$ value using the measure as a predictor, and the Nugent score as response in a linear model.
  Amsel accuracy: proportion of specimens with correct BV classification under a leave-one-out cross-validation.
  Amsel p-value: p-value from a two-sample $t$-test on values stratified by BV classification.
  ``OTU'' designates the measure applied to 97\% clustering groups, and ``Family'' designates taxonomic classification at the family level.
  Measures described in main text.
}
\label{TABLEvaginalcorrelation}
\end{table}}

\newcommand{\FIGoralboxplots}{\
  \begin{figure}
    \begin{center}
      \centerline{\
      \includegraphics[width=13cm]{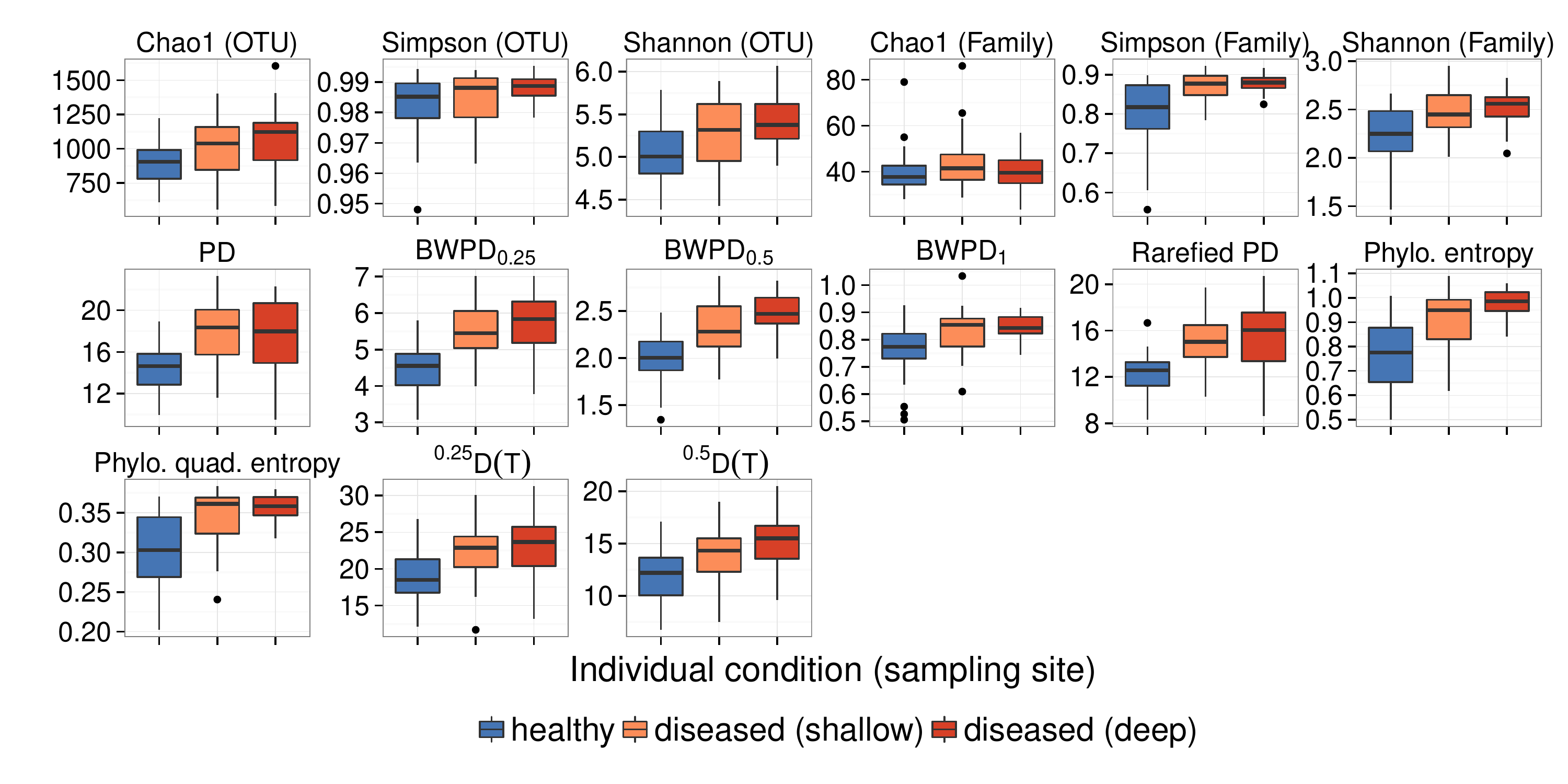}}
    \end{center}
    \caption{Comparison of diversity between samples from healthy controls, healthy sites of diseased patients, and diseased sites of diseased patients in the oral dataset, using different measures of alpha diversity.
    Top row: cluster-based methods.
    Bottom row: phylogenetic methods.}
    \label{FIGoralboxplots}
  \end{figure}
}

\newcommand{\TABLEskinANOVA}{\
\begin{table}[ht]
\centering
\begin{tabular}{rrrrrr}
  \hline
   & Ac & N & Pc & Vf & mean rank \\
  \hline
  \BWPDa{0.25} & 2.90e-02 & 1.29e-03 & 4.94e-03 & 1.71e-04 & 3.50 \\
  \PD & 2.72e-02 & 5.48e-03 & 8.62e-03 & 5.06e-04 & 5.25 \\
  \qDT{0.25} & 2.95e-02 & 1.24e-03 & 1.53e-02 & 3.86e-04 & 5.50 \\
  \qDT{0.5} & 3.03e-02 & 4.95e-04 & 2.92e-02 & 3.20e-04 & 5.75 \\
  \BWPDa{0.5} & 6.53e-02 & 7.34e-05 & 1.43e-02 & 1.42e-03 & 6.75 \\
  \qDT{0} & 3.08e-02 & 2.37e-03 & 9.77e-03 & 8.88e-04 & 7.00 \\
  Chao1 (OTU) & 2.97e-02 & 2.68e-03 & 9.14e-03 & 9.97e-03 & 7.00 \\
  Shannon (OTU) & 7.09e-02 & 8.48e-02 & 1.23e-01 & 2.70e-05 & 8.00 \\
  Phylo. entropy & 1.17e-01 & 2.31e-05 & 8.37e-02 & 1.03e-02 & 8.25 \\
  Phylo. quad. entropy & 2.52e-01 & 6.31e-06 & 4.77e-01 & 1.55e-01 & 9.00 \\
  Simpson (OTU) & 1.17e-01 & 3.68e-01 & 8.75e-01 & 1.15e-04 & 9.50 \\
  \BWPDa{1} & 3.11e-01 & 2.99e-05 & 6.45e-01 & 5.33e-01 & 10.25 \\
  \hline
\end{tabular}
\caption{ANOVA p-values for various phylogenetic diversity statistics applied to the skin microbiome data of \citet{oh2012shifts}.
Rows are ordered by increasing mean rank across sites.
The same site abbreviations are used as in their paper: Af, antecubital fossa; N, nares; Pf, popliteal fossa; Vf, volar forearm.
}
\label{TABLEskinANOVA}
\end{table}
}
\newcommand{\refTABLEskinANOVA}{5}

\begin{document}

% PeerJ: max 200 chars
% For inline equations use the \(...\) notation.
% For display equations use the $$...$$ or \[...\] notation.
\title[Abundance-weighted phylogenetic diversity measures]{Abundance-weighted phylogenetic diversity measures distinguish microbial community states and are robust to sampling depth}
\author{Connor O. McCoy}
\author{Frederick A. Matsen IV}
\date{\today}

\arxiv{\
\begin{abstract}
% PeerJ: max 3000 chars
    In microbial ecology studies, the most commonly used ways of investigating alpha (within-sample) diversity are either to apply count-only measures such as Simpson's index to Operational Taxonomic Unit (OTU) groupings, or to use classical phylogenetic diversity (PD), which is not abundance-weighted.
    Although alpha diversity measures that use abundance information in a phylogenetic framework do exist, but are not widely used within the microbial ecology community.
    The performance of abundance-weighted phylogenetic diversity measures compared to classical discrete measures has not been explored, and the behavior of these measures under rarefaction (sub-sampling) is not yet clear.
    In this paper we compare the ability of various alpha diversity measures to distinguish between different community states in the human microbiome for three different data sets.
    We also present and compare a novel one-parameter family of alpha diversity measures, \(\operatorname{BWPD}_\theta\), that interpolates between classical phylogenetic diversity (PD) and an abundance-weighted extension of PD\@.
    Additionally, we examine the sensitivity of these phylogenetic diversity measures to sampling, via computational experiments and by deriving a closed form solution for the expectation of phylogenetic quadratic entropy under re-sampling.
    In all three of the datasets considered, an abundance-weighted measure is the best differentiator between community states.
    OTU-based measures, on the other hand, are less effective in distinguishing community types.
    In addition, abundance-weighted phylogenetic diversity measures are less sensitive to differing sampling intensity than their unweighted counterparts.
    Based on these results we encourage the use of abundance-weighted phylogenetic diversity measures, especially for cases such as microbial ecology where species delimitation is difficult.
\end{abstract}
}

% \section{Keywords}
% alpha diversity; phylogenetic diversity; diversity indices

\maketitle

%\noindent
%Connor O. McCoy and Frederick A. Matsen IV$^*$\\
%Fred Hutchinson Cancer Research Center\\
%1100 Fairvew Ave. N\\
%Seattle, WA 98109\\
%
%\noindent
%$^*$Corresponding author: matsen@fhcrc.org
%\newpage

\section{Introduction}

It is now well accepted that incorporating phylogenetic information into alpha (single-sample) and beta (between-sample) diversity measures can be useful in a variety of ecological contexts.
Phylogenetic equivalents of all of major alpha diversity measures have been developed.
Starting with Faith's original definition of phylogenetic diversity \citep{faith1992conservation}, which generalizes species count, there are now phylogenetic generalizations of the Simpson index to Rao's quadratic entropy \citep{rao1982diversity,warwick1995new}, the Shannon index to phylogenetic entropy \citep{allen2009new}, and the Hill numbers to \qDT{q} \citep{chao2010phylogenetic}.
Phylogenetic diversity itself has been extended to incorporate taxon counts \citep{barker2002phylogenetic} and proportional abundance \citep{vellend2011measuring}.
There have also been abundance-weighted measures that explicitly measure phylogenetic community structure \citep{fine2011phylogenetic}, or an ``effective number of species'' \citep{chao2010phylogenetic}.
Many diversity measures can be tidily expressed in the framework of \citet{leinster2012measuring}, although the expression of phylogenetic diversity measures for non-ultrametric trees is complex.

In this paper we use three example human microbiome datasets to demonstrate the utility of abundance-weighted phylogenetic diversity measures.
We also introduce a one-parameter family interpolating between classical PD and an abundance-weighted generalization.
We call the parameter $\theta$ and denote the one-parameter family $\BWPDt$; $\BWPDa{0}$ is classical PD, whereas $\BWPDa{1}$ is balance-weighted phylogenetic diversity, effectively \PDaw\ of \citet{vellend2011measuring}.
Intermediate values of $\theta$ allow a partially-abundance-weighted compromise.
Such a compromise has recently been shown to be useful for measuring beta diversity, with the introduction of a one-parameter family of ``generalized UniFrac'' measures \citep{chen2012associating}.
We use the name Balance Weighted Phylogenetic Diversity as described below because there are a variety of abundance weighted phylogenetic diversity measures.
We compare the behavior of PD measures, including \BWPDt, under various levels of sampling using theory and example data sets.

\section{Materials and methods}

\subsection{Datasets}

We apply the methods described below to three previously described 16S rRNA surveys of the human microbiome.
The first two datasets are composed of samples from ``normal'' and dysbiotic microbial communities, where previous studies have associated changes in diversity with changes in health.
The third dataset investigates the changes of the skin microbiome through time.

\subsubsection{Bacterial vaginosis}

First, we reanalyze a pyrosequencing dataset describing bacterial communities from women being monitored in a sexually transmitted disease clinic for bacterial vaginosis (BV).
BV has previously been shown to be associated with increased community diversity \citep{fredricks2005molecular}.
For this study, swabs were taken from 242 women from the Public Health, Seattle and King County Sexually Transmitted Diseases Clinic between September 2006 and June 2010 of which 220 samples resulted in enough material to analyze \citep{srinivasan2012bacterial}.

Selection of reference sequences and sequence preprocessing were performed using the methods described in \citep{srinivasan2012bacterial}.
452,358 reads passed quality filtering, with a median of 1,779 reads per sample (range: 523--2,366).

\subsubsection{Oral periodontitis}

We also utilize sequence data from a study of subgingival communities in 29 subjects with periodontitis, along with an equal number of healthy controls \citep{griffen2011distinct}.
The publication analyzing this dataset showed increased community diversity in samples from diseased patients compared to healthy controls.
Raw sequences were filtered, retaining only those reads with: a mean quality score of at least 25, no ambiguous bases, at least 150 base pairs in length, and an exact match to the sequencing primer and barcode.
A total of 759,423 reads passed quality filtering, with a median of 8,320 reads per sample (range: 4,096--14,319).

As the phylogenetic placement method used below to calculate our measures requires a reference tree and alignment, we created a tree with FastTree 2.1.4 \citep{price2010fasttree} using the alignment and accompanying taxonomic annotation from the curated CORE database of oral microbiota \citep{griffen2011core}.

\subsubsection{Skin microbiome through time}

Our third data set is a study of skin microbial diversity through adolescence \cite{oh2012shifts}.
Aligned sequences were obtained directly from the authors, although sequence data is available under the accession numbers [GQ000001] to [GQ116391] and can be accessed through BioProject ID 46333.

\subsection{Balance-weighted phylogenetic diversity}

In this section we introduce \BWPDt, our one-parameter family interpolating between classical PD and fully balance-weighted phylogenetic diversity.
We will primarily consider so-called \emph{unrooted} \citep{pardi2007resource} phylogenetic diversity, which does not necessarily include the root.
The case of \emph{rooted} phylogenetic diversity can be calculated in a similar though simpler way as described below.
Although we will primarily be working in an unrooted sense, it will be useful to use terminology that corresponds to the rooted case.
For this reason, if the tree is not already rooted, assume an arbitrary root has been chosen; let the \emph{proximal} side of a given edge be the side that contains the root and \emph{distal} be the other.

We will describe \BWPDt\ in terms of a phylogenetic tree $T$ with leaves $L$, and a \emph{contingency table} describing the number of observations of the organisms at the leaves in various samples.
The contingency table has rows labeled with the leaves of $T$, and columns labeled by samples.
In microbial ecology this is frequently known as an \emph{OTU table}.
The entry corresponding to a given leaf and a given sample is the number of times that leaf was observed in that sample.

The classical (unrooted) phylogenetic diversity of a given sample in this context is the total branch length of the tree subtended by the leaves in that sample.

The path to generalizing PD is to note that this can be expressed as a sum of branch lengths multiplied by a step function.
Let $f(x)$ be the function that is one for $x > 0$ and zero otherwise.
Let $g(x) = \min(f(x),f(1-x))$ and $D_s(i)$ be the fraction of reads in sample $s$ that are in leaves on the distal side of edge $i$.
Phylogenetic diversity can be then expressed as
\begin{equation}
  \PD(s) = \sum_i \ell_i \, g (D_s(i))
  \label{eqn:PDdef}
\end{equation}
That is, the sum of edge lengths in $T$ which have reads from $s$ on both the distal and proximal side.

\arxiv{\FIGshape}

Note that the step function $g$ is the limit of a one-parameter family of functions (Fig.~\ref{FIGshape}).
Indeed, defining
\begin{equation}
  g_\theta(x) = \min \left( x^\theta, (1-x)^\theta \right),
\end{equation}
$g$ is the pointwise limit on the closed unit interval of the $g_\theta$ as $\theta$ goes to zero.
Thus our one-parameter generalization is
\begin{equation}
  \BWPDt(s) = \sum_i \ell_i \, g_\theta(D_s(i)).
  \label{eqn:expdPDdef}
\end{equation}
Note that when $\theta=0$ this is PD and when $\theta=1$ this is an abundance-weighted version of PD equivalent to executing the \dnPD\ recipe of \citet{barker2002phylogenetic} up to a multiplicative factor.

The rooted equivalent of \eqref{eqn:expdPDdef} is
\begin{equation}
  \RBWPDt(s) = \sum_i \ell_i \, (D_s(i))^\theta,
  \label{eqn:expdPDdefRooted}
\end{equation}
which interpolates between rooted PD and an abundance-weighted version.
\citet{vellend2011measuring} describe a similar measure, \PDaw, which is equal to \RBWPDa{1} multiplied by the total number of branches in $T$.

We call \BWPD\ balance-weighted phylogenetic diversity because it weights edges according to the balance of read fractions on either side of an edge-- edges with even amount of mass on either side are up-weighted, while edges with an uneven balance of mass are down-weighted.
Indeed, if $|x-(1-x)|$ is taken to represent the imbalance of read fraction on either side of an edge, then $1-|x-(1-x)|$ can be taken to be a measure of balance; note that on the unit interval, $\min(x,1-x) = 1-|x-(1-x)|$.
Because a small $x$ or an $x$ close to 1 gives a small coefficient in the summation, small collections of reads or small perturbations of the read distribution will not change the value of \BWPD\ appreciably.

\subsection{Calculation of PD measures in example applications}

Reads from the vaginal and oral studies were placed on a tree created from a curated set of taxonomically annotated reference sequences.
As phylogenetic entropy and \qDT{q} operate on a rooted phylogeny, reference trees were assigned a root taxonomically \citep{matsen2012reconciling}.
\pplacer\ was run in posterior probability mode (using the \verb|-p| and \verb|--informative-prior| flags), which defines an informative prior for pendant branch lengths with a mean derived from the average distances from the edge in question to the leaves of the tree.
The resulting set of placements were classified at the family rank using a hybrid classifier implemented in the \guppy\ tool from the \pplacer\ suite.
The hybrid classifier assigns taxonomic annotations to sequences using the combination of a na\"{i}ve Bayes classifier \citep{wang2007naive} with a phylogenetic classifier (Matsen et al., unpublished results).
Any reads that could not be confidently classified to the family rank were not used in measures based on classification.

Full-length 16S sequences were available for the skin data, and so a more traditional tree-building approach was used.
Representative OTUs were chosen for each site by clustering at 97\% identity using USEARCH \citep{edgar2010search}, with trees built on OTU centroids using FastTree \citep{price2010fasttree}.
To conform with methods used in that paper, the na\"ive Bayes classifier \citep{wang2007naive} was used to infer genus-level classifications to taxonomically root the tree; in our case we used the RDP classifier v2.5.
The contingency (OTU) tables generated by clustering were made available to our tools via the BIOM \citep{mcdonald2012biological} format.

\PD\ (unrooted PD), phylogenetic quadratic entropy \citep{rao1982diversity}, phylogenetic entropy \citep{allen2009new}, and $\qDT{q}$ \citep{chao2010phylogenetic} were implemented for phylogenetic placements in the freely-available \pplacer\ suite of tools \citep{matsenEaPplacer10} (\url{http://matsen.fhcrc.org/pplacer}) in the subcommand \textsf{guppy fpd}.
\PD\ on rarefied phylogenetic placements was calculated using \textsf{guppy rarefy}.

Discrete measures of alpha diversity and richness were calculated on contingency tables obtained from clustering and taxonomic classification.
Sequences were clustered into Operational Taxonomic Units (OTUs) at a 97\% identity threshold using \USEARCH\ 5.1 \citep{edgar2010search}.
Similar results were observed when clustering at 95\% identity (results not shown).
OTU counts and family-level taxon counts were then rarefied to the read count of the specimen in the dataset with the fewest sequences in \R\ 2.15.1 \citep{rteam2012r} using the \vegan\ package \citep{oksanen2012vegan}.
We obtained values for the \citet{simpson1949measurement} and \citet{shannon1948mathematical} diversity indices, as well as the Chao1 \citep{chao1984nonparametric} and ACE \citep{chao1992estimating} measures of species richness using \vegan\ functions \textsf{diversity} and \textsf{estimateR}.

\subsection{Comparative analysis of alpha diversity measures}

To investigate the relation between various measures of alpha diversity, we calculated Pearson's $r$ between all pairs of measures using the function \textsf{rcorr} from the \R\ package \Hmisc\ \citep{harrell2012hmisc}.
We then performed hierarchical clustering with the \R\ function \textsf{hclust}, using $d = 1 - r$ as the distance between two measures.

Association of each measure with clinical criteria for the first two data sets was evaluated by examining the accuracy of a logistic regression using the measure as the sole predictor of whether the sample came from a ``normal'' or dysbiotic subject.
In the vaginal dataset, we assessed each measure's ability to predict whether a sample was from a subject positive for BV by Amsel's criteria, a clinical diagnostic method \citep{amsel1983nonspecific}.
In the oral dataset, we assessed each measure's ability to predict whether a sample was from a healthy control, or a subject with periodontitis.
Accuracy in predicting sample community state was assessed by leave-one-out cross-validation using the \R\ package \boot\ \citep{davison1997bootstrap,canty2012boot}.

For the vaginal dataset, we also calculated $R^2$ values using each measure individually as a predictor for sample Nugent score in a linear regression.
The Nugent score provides a diagnostic score for BV which ranges from 0 (BV-negative) to 10 (BV-positive) based on presence and absence of bacterial morphotypes as viewed under a microscope \citep{nugent1991reliability}.

We calculated p-values to compare within- and between-stratification variability using R's built-in \textsf{t.test} function for the vaginal data, which had a binary stratification, and \textsf{aov} function for the oral and vaginal data sets.
The vaginal dataset data was stratified by Amsel's criterion, the oral dataset by condition and sampling site, and the skin microbiome dataset by Tanner scale of physical development \citep{oh2012shifts}.

Plots were prepared with \R\ base graphics and \ggplot\ \citep{wickham2009ggplot2}.

\subsection{Evaluation of performance under rarefaction}

Phylogenetic placements were rarefied using the \textsf{rarefy} subcommand of the \textsf{guppy} tool in the \pplacer\ suite.
Phylogenetic alpha diversity measures were calculated on the resulting rarefied placements as described above.

\section{Results}

\subsection{Application to the human microbiome}

\subsubsection{Vaginal microbiome}

Like \citet{srinivasan2012bacterial} and many others in the field, we observe greater diversity in BV positive specimens using a variety of diversity and richness measures (Fig.~\ref{FIGvaginalboxplots}).
In particular, this is true for \BWPDt\ for a variety of values of $\theta$ (Fig.~\ref{FIGvaginalnugkappa}).

\arxiv{\TABLEvaginalcorrelation}
In the vaginal data, phylogenetic measures of alpha diversity have better cross-validation accuracy for the Amsel classification and better correlation with the Nugent score than discrete OTU-based measures (Table~\ref{TABLEvaginalcorrelation}).
All measures were somewhat accurate in identifying community state, with even the worst performers classifying at least 70\% of samples correctly.
\BWPDa{0.25}, rarefied \PD, \PD, and phylogenetic entropy perform equally well predicting BV status.
Correlation with Nugent score varies more widely, from 0.19 using Simpson (OTU) to 0.74 using \BWPDa{0.25} or Simpson applied to family-level classifications.
OTU-based measures rank in the bottom half of the measures tested, and below all phylogenetic measures.
Phylogenetic diversity, which can be viewed as a measure of richness, outperforms discrete measures of richness, and most measures incorporating abundance.

\arxiv{\FIGvaginaldendro}
In the hierarchical clustering of alpha measures on the vaginal data set, phylogenetic methods are separated from OTU-based methods (Fig.~\ref{FIGvaginaldendro}).
\BWPDt\ is similar to different extant phylogenetic alpha diversity measures for different $\theta$.
The Simpson and Shannon diversity measures cluster together, as do the ACE and Chao1 richness measures.

\arxiv{\FIGvaginalrarefy}
Fig.~\ref{FIGvaginalrarefy}\ shows values of \BWPDt\ calculated before ($x$-axis) and after ($y$-axis) a single rarefaction to 523 sequences per sample.
Samples for which the \BWPDt\ value changes little lie close to the blue line, which shows the case of no difference between original and rarefied samples.
Increasing $\theta$, which corresponds to increased use of abundance information, reduces the change in \BWPDt\ induced by rarefaction.
Phylogenetic quadratic entropy and phylogenetic entropy both show behavior similar to \BWPDa{1}, with rarefaction introducing little effect.

It might be possible to formalize a statement to this effect by computing the expectation of these alpha measures under rarefaction.
However, computing the expectation for \BWPDt\ under rarefaction does not appear to be straightforward: the methods of \citet{dremin1994fractional} might be applicable in this setting, however, even the integer moments of the hypergeometric distribution are complicated and the non-integer moments are bound to be very complex.
We have, however, shown in the Appendix that the expectation of phylogenetic quadratic entropy under rarefaction to $k$ sequences assigned to the tips of a phylogenetic tree is
\[
  \EE[\PQE_k] = \frac{k-1}{k n (n-1)} \sum_i \ell_i d_i (n-d_i)
\]
where $d_i$ is the number of sequences falling below edge $i$ and $\ell_i$ is the length of edge $i$.
This is almost identical to the unrarefied value of phylogenetic quadratic entropy, i.e.\
\[
  \PQE = \frac{1}{n^2} \sum_i \ell_i d_i (n-d_i).
\]
Thus it is not surprising to see that the expectation of PQE under rarefaction is very close to the original value (Fig.~\ref{FIGvaginalepqe}) for reasonably large $k$ and $n$.

\subsubsection{Oral microbiome}

As previously observed by \citet{griffen2011distinct}, we find generally higher diversity in samples from diseased patients (Fig.~\ref{FIGoralboxplots}).
We evaluated the ability of each alpha diversity measure to predict whether a sample came from an individual with periodontitis, regardless of sample collection site, using the above methods.

In the oral dataset, phylogenetic alpha diversity measures incorporating abundance gave the best predictions of community state (Table~\ref{TABLEoralaccuracy}, Fig.~\ref{FIGoralboxplots}).
In contrast, classical phylogenetic diversity was amongst the worst predictors; rarefaction did help, but rarefied PD still performed worse than phylogenetic measures taking abundance into account.

OTU-based methods and phylogenetic methods are not as separated in a hierarchical clustering as for the vaginal dataset (Fig.~\ref{FIGoraldendro}).
However, many of the same pairings are present in both clusterings: \BWPDa{0.5}\ with PE, \BWPDa{1}\ with QE, Simpson with Shannon, ACE with Chao1, and \PD\ with rarefied PD\@.
Interestingly, \PD, rarefied \PD, and \BWPDa{0.25} all cluster with the discrete richness measures ACE and Chao1.

Like the vaginal dataset, incorporating abundance information decreases the effect of rarefaction on \BWPDt\ values (Figs.~\ref{FIGoralrarefy},~\ref{FIGoralepqe}).

\arxiv{\FIGoralboxplots}

\subsubsection{Skin microbiome}

To further assess resolution and robustness of abundance weighted phylogenetic diversity measures, we considered skin microbiome data from a study by \citet{oh2012shifts}.
This study tracked the changes of the skin microbiome through developmental stages.
Because there are five Tanner stages, and they do not have a monotonic relationship with skin microbiome diversity \citep{oh2012shifts}, we focused on ANOVA p-values to see if the diversity measurements had small within-stage heterogeneity compared to between-stage heterogeneity.
To compare the ANOVA p-values associated with the diversity measurements across the various data sets, we ranked the p-value of the diversity measures from lowest to highest for each data set individually.
We averaged these ranks to gain an overall measure of performance.
The results again show phylogenetic measures generally performing better than OTU-based measures (Tab.~\refTABLEskinANOVA).
In this case, a light weighting or no weighting of phylogenetic diversity by abundance performed better than full abundance-weighting.
Note that we are not presenting these uncorrected p-values as evidence that there is an interesting relationship between skin microbiome and developmental stage, but rather are using p-values as a way of measuring within-stage heterogeneity compared to between-stage heterogeneity for the various measures.

\TABLEskinANOVA

\subsubsection{Applications summary}

In all three of the data sets investigated, abundance-weighted phylogenetic diversity measures showed good performance to distinguish between community states: between ``normal'' and dysbiotic samples in the oral and vaginal microbiomes, and between developmental stages in the skin microbiome.
Notably, the best distinguishing measure in each dataset was both phylogenetic and abundance-weighted.
\BWPDt, our new family of abundance-weighted phylogenetic diversity measures, was highly correlated with clinical status although the value of $\theta$ most associated with community state varied.
On the vaginal and oral data sets intermediate values of $\theta$ for $\BWPDt$ provide the best correlation with clinical status.
These results correspond to analogous results for beta diversity, where an intermediate exponent for ``generalized UniFrac'' was the most powerful \citep{chen2012associating}.

\section{Discussion}

Phylogenetic alpha diversity measures were more closely related to community state than were discrete measures based on OTU clustering for the data sets investigated here.
This result is especially interesting given that the Simpson index, the Shannon index, or counting applied to OTU tables are very common ways of characterizing microbial diversity \citep{fierer2007metagenomic,grice2009topographical,hill2003using,dethlefsen2011incomplete}.
As also noted by \citet{aagaard2012metagenomic}, we find that measurements of diversity using taxonomic classification can be useful in describing communities, and in fact perform much better than the same measurements of diversity applied to OTU counts; however, this approach requires a taxonomically well characterized environment.
Our results can be viewed as an experimental confirmation of the notion that incorporating similarity between species is important to get sensible measures of diversity, which has been advocated by many, including most recently by \citet{leinster2012measuring}.

% Do we want to speculate why it might be so-- i.e. we need to take the phylogenetic relationships into account?
% @Connor: Well, I guess I think of it as being related to the fact that with OTUs you only get one cutoff value, but with phylo you can get a whole range in a sense. Thoughts?
% @Erick I think of it the same way.

% @Erick: we don't actually show any data on rarefied vs. unrarefied OTU tables here, hence the change.
We find that classical phylogenetic diversity is sensitive to sampling depth, underestimating the true value in small samples.
Biases have also been described for diversity measures using OTU tables \citep{gihring2012massively}.
In contrast, we observe that some abundance-weighted phylogenetic measures are relatively robust to varying levels of sampling.

As of the publication of this paper, no abundance-weighted phylogenetic alpha diversity measures are implemented in either mothur \citep{schloss2009introducing} or QIIME \citep{caporaso2010qiime}, two of the most popular tools for analysis of microbial ecology data.
Although the fact that abundance-weighted phylogenetic diversity measures performed best for the three data sets investigated here does not imply that they are best in general, we suggest that abundance-weighted phylogenetic measures be given greater consideration for microbial ecology studies.
For this to happen, implementations in commonly used microbial ecology software packages will be needed, in addition to our implementation and that of the \picante\ R package \citep{kembel2010picante}.

\section{Acknowledgements}
The authors would like to thank Steven Kembel for encouragement and guidance, Steven N. Evans for probability consultation, and David Nipperess for an interesting dialog concerning phylogenetic diversity and rarefaction.
The Segre lab at the NIH, in particular Sean Conlan and Julia Oh, were very generous and helpful with the skin data.
This work would not have been possible without an ongoing collaboration with David Fredricks, Noah Hoffman, Martin Morgan, and Sujatha Srinivasan at the Fred Hutchinson Cancer Research Center.
This work was supported in part by NIH R01 HG005966-01.

\bibliography{alpha}
\bibliographystyle{alphaplainnat}

% All figures, for submission
\notarxiv{\newpage\FIGshape}
\notarxiv{\newpage\TABLEvaginalcorrelation}
\notarxiv{\newpage\FIGvaginaldendro}
\notarxiv{\newpage\FIGvaginalrarefy}
\notarxiv{\newpage\FIGoralboxplots}

\end{document}

\section{Appendix}

\subsection{Rarefaction of phylogenetic quadratic entropy}

We investigate the rarefaction of phylogenetic quadratic entropy (PQE), which is a diversity coefficient in the language of \citep{rao1982diversity} and defined as follows \citep{warwick1995new,allen2009new}.
If the tree is not rooted, root it arbitrarily (the rooting does not impact the value).
PQE is defined on a tree as
\begin{equation}
  \PQE_k = \sum_i \ell_i \frac{a_i}{n} \left( 1-\frac{a_i}{n} \right),
  \label{eq:PQE}
\end{equation}
where $\ell_i$ is the length of edge $i$ and $a_i$ is the number of leaf observations that are distal (away from root) from edge $i$.

Assume we rarefy to $k$ observations as above; let $A_i$ denote the random variable that is the number of observations distal to edge $i$ after rarefaction.
The phylogenetic quadratic entropy is then
\begin{equation}
  \PQE_k = \sum_i \ell_i \frac{A_i}{k} \left( 1-\frac{A_i}{k} \right).
\end{equation}

The random variable $A_i$ has a hypergeometric distribution, performing $k$ draws with $d_i$ possible successes in a population of size $n$.
Let $\mu_i$ be the expectation of $A_i$, which is simply $k d_i/n$.
The variance of the hypergeometric distribution is well known to be
\begin{equation}
  \sigma_i^2 = \frac{k d_i (n-d_i) (n-k)}{n^2 (n-1)}.
\end{equation}

% Compare Wikipedia:
% For a hypergeometric distribution with k draws from a population of size n with d successes:
% mean = k r
%              d  (n-d) (n-k)
% sigma^2 = k --- ----- -----
%              n    n   (n-1)

Next
\[
\begin{split}
  \EE[\PQE_k] & = \sum_i \ell_i \EE \left[ \frac{A_i}{k} \left( 1-\frac{A_i}{k} \right) \right], \\
              & = \frac{1}{k^2} \sum_i \ell_i \left(k \EE[A_i]-\EE[A_i^2] \right).
\end{split}
\]

By definition,
\[
  \EE[A_i^2] = \mu_i^2 + \sigma_i^2.
\]
Thus the expectation of the phylogenetic quadratic entropy upon rarefaction is
\begin{equation}
  \EE[\PQE_k] = \frac{1}{k^2} \sum_i \ell_i (k \mu_i - \mu_i^2 - \sigma_i^2).
  \label{eq:PQEexprarefact}
\end{equation}

Expanding the term in parentheses from \eqref{eq:PQEexprarefact}:
\[
  \begin{split}
    k \mu_i - \mu_i^2 - \sigma_i^2 & = k^2 \frac{d_i}{n} - k^2 \frac{d_i^2}{n^2} - \frac{k d_i (n-d_i) (n-k)}{n^2 (n-1)} \\
                                   % & = \frac{k^2 d_i n (n-1) - k^2 d_i^2 (n-1) - k d_i (n-d_i) (n-k)}{n^2 (n-1)} \\
                                   & = k d_i \frac{k n (n-1) - k d_i (n-1) - (n-d_i) (n-k)}{n^2 (n-1)} \\
                                   % & = k d_i \frac{(n-d_i) k (n-1) - (n-d_i) (n-k)}{n^2 (n-1)} \\
                                   & = k d_i \frac{(n-d_i) (k (n-1) - (n-k))}{n^2 (n-1)} \\
                                   % & = k d_i \frac{(n-d_i) (kn-k - n+k))}{n^2 (n-1)} \\
                                   % & = k d_i \frac{(n-d_i) n (k-1))}{n^2 (n-1)} \\
                                   & = \frac{k (k-1)}{n (n-1)} d_i (n-d_i) \\
  \end{split}
\]

Putting this back in \eqref{eq:PQEexprarefact}, we obtain
\begin{equation}
  \EE[\PQE_k] = \frac{k-1}{k n (n-1)} \sum_i \ell_i d_i (n-d_i)
\end{equation}

In principle one could calculate the variance of phylogenetic quadratic entropy in terms of the higher order moments of the hypergeometric distribution.
However, these higher moments are very messy and we have not attempted to write out the variance calculation.
We also note that this derivation could be easily generalized to the setting of a ``tree with marks'' as in \citep{nipperess2012mean}.

\subsection{Description of PD in more general setting}

We can describe the methods in the general setting where samples are represented by a mass distribution on a tree.
As described elsewhere \citep{EvansMatsenPhyloKR12}, this generalizes the notion of representing a sample by an OTU count equipped with a phylogenetic tree on OTU representative.
Specifically, if the total sample size is $N$, then $n$ observations of a given OTU $\omega$ are represented by a point mass of weight $n/N$ at $\omega$.

As observed by others \citep{allen2009new} phylogenetic diversity measures can be written as
\begin{equation}
  \PD(s) = \sum_{i} \ell_i F(D_s(i))
\end{equation}
where $F$ is some real-valued function on the unit interval.
This can be further generalized to the case of an abitrary probability distribution by writing this as an integral
where $\lambda$ is the length measure on the tree \citep{EvansMatsenPhyloKR12} and now $D_s(y)$ is the total mass on the distal side of $y$.
\begin{equation}
  \PD(s) = \int_{y \in T} F(D_s(y)) \, \lambda(dy)
\end{equation}
For phylogenetic quadratic diversity, $F(x) = x(1-x)$, and for phylogenetic entropy, $F(x)= - x \log x$.
As described above, the \BWPDt\ fits into this framework with $F(x) = \min(g_\theta(x),g_\theta(1-x))$.

\bibliography{alpha}
\bibliographystyle{alphaplainnat}

\newpage
\notarxiv{\newpage}\FIGvaginalboxplots
\notarxiv{\newpage}\FIGvaginalnugkappa
\notarxiv{\newpage}\FIGvaginalepqe
%\FIGvaginalfattree
\notarxiv{\newpage}\FIGoralrarefy
\notarxiv{\newpage}\FIGoralepqe
\notarxiv{\newpage}\FIGoraldendro
%\FIGoralfattree
\notarxiv{\newpage}\TABLEoralaccuracy